\documentclass[preprintnumbers]{revtex4}

\usepackage{graphicx}

\begin{document}

\preprint{P5\_sullivan\_0709}
\preprint{ANL-HEP-CP-01-104}
\preprint{FERMILAB-Conf-01/329-T}
\preprint{MSUHEP-11026}
\preprint{SMU-HEP 01/12}

\title{Heavy-quark parton distribution functions and their uncertainties}

\author{Z. Sullivan\protect$ ^{1,2}$}
\email{zack@fnal.gov}
\thanks{work at Argonne supported by the U. S. Department of Energy 
under contract No. W-31-109-ENG-38}
\thanks{work at Fermilab supported by the U. S. Department of Energy 
under contract No. DE-AC02-76CH03000}
\thanks{work at Snowmass supported in part by an American Physical Society 
Division of Particles and Fields Snowmass 2001 Fellowship.}
\affiliation{$^{1} $High Energy Physics Division, Argonne National
Laboratory, Argonne, IL 60439 \\
$^{2} $Theoretical Physics Department, Fermi National Accelerator
Laboratory, Batavia, Illinois 60510-0500}

\author{P. M. Nadolsky\protect$ ^{3,4}$}
\email{nadolsky@pa.msu.edu}
\thanks{work supported by the National Science Foundation, U. S. Department
of Energy, and Lightner-Sams Foundation.}
\affiliation{$^{3} $Department of Physics \& Astronomy, Michigan State 
University, East Lansing, MI, 48824 \\
$^{4} $Department of Physics, Southern Methodist University, Dallas,
TX, 75275}

\date{November 20, 2001}

\begin{abstract}
We investigate the uncertainties of the heavy-quark parton
distribution functions in the variable flavor number scheme. Because
the charm- and bottom-quark parton distribution functions (PDFs) are
constructed predominantly from the gluon PDF, it is a common practice
to assume that the heavy-quark and gluon uncertainties are the
same. We show that this approximation is a reasonable first guess, but
it is better for bottom quarks than charm quarks. We calculate the PDF
uncertainty for $t$-channel single-top-quark production using the
Hessian matrix method, and predict a cross section of
$2.12^{+0.32}_{-0.29}$~pb at run II of the Tevatron.
\end{abstract}
\maketitle

\vspace*{-1em}
As a new run of the Fermilab Tevatron begins, there is a considerable
interest in measuring fundamental parameters of the Standard Model and
in looking for new particles. For some of these measurements the
dominant uncertainty will come from parton distribution functions
(PDFs). In particular, measurements of the Cabibbo-Kobayashi-Maskwawa
(CKM) matrix element $V_{tb}$ depend on an accurate prediction of the
cross section for single-top-quark production \cite{Stelzer:1997ns}
and will be limited by uncertainties in the $b$-quark PDF
\cite{Stelzer:1998ni}. In a large $\tan \beta$ model of supersymmetry,
Higgs production from a $b\bar{b}$ initial state is the dominant
production mechanism \cite{Dicus:1998hs}.

In order to understand uncertainties of the heavy-quark parton
distribution functions, it is necessary to remember where they
arise. In the conventional QCD framework, heavy quarks are produced
perturbatively from gluons splitting into heavy-quark pairs at
energies above the heavy-quark mass threshold (we do not consider
``intrinsic'' heavy quarks \cite{Brodsky:1980pb} here).  The
heavy-quark PDF $Q(x,\mu^2)$ is a formal object, which can be used
reliably when the typical momentum scale $\mu$ of the hard scattering
is much larger than the mass of the heavy quark $m_Q$. This PDF resums
nearly collinear singularities $\alpha_s^n\ln^n(\mu^2/m_Q^2)/n!$
associated with propagation of the initial-state heavy quarks. Unlike
the light-quark PDFs, the distributions for charm ($c$) and bottom
quarks ($b$) are not independent functions that are determined by
fitting to the hadronic data; instead, they are generated from the
PDFs of light partons in the process of
Dokshitzer-Gribov-Lipatov-Altarelli-Parisi (DGLAP) evolution.

The relationship between the PDFs for heavy and light partons is
especially simple near the mass threshold of heavy quarks ($\mu\sim
m_Q$), where the DGLAP equations can be solved in the
leading-logarithm approximation.  Given the initial condition
$Q(x,\mu^2)=0$ at $\mu=m_Q$, and neglecting gluon bremsstrahlung off
heavy-quark lines and the scale dependence of the gluon PDF and
$\alpha_s$, the leading-order solution can be found as
[5--7]
\begin{equation}
\label{P5_sullivan_0709eq_b}
Q(x,\mu^2) = \frac{\alpha_s(\mu^2)}{2\pi}\ln 
\left( \frac{\mu^2}{m_Q^2}\right) 
\int_{x}^{1}\frac{dz}{z}P_{Qg}\left( \frac{x}{z}\right) 
g\left( z,\mu^2\right) \,,
\end{equation}
where the function $P_{Qg}(y)\equiv \left[y^2+(1-y)^2\right] /2$
describes the splitting of a gluon into a heavy-quark pair.

\begin{figure}[tb]
\begin{center}
\includegraphics{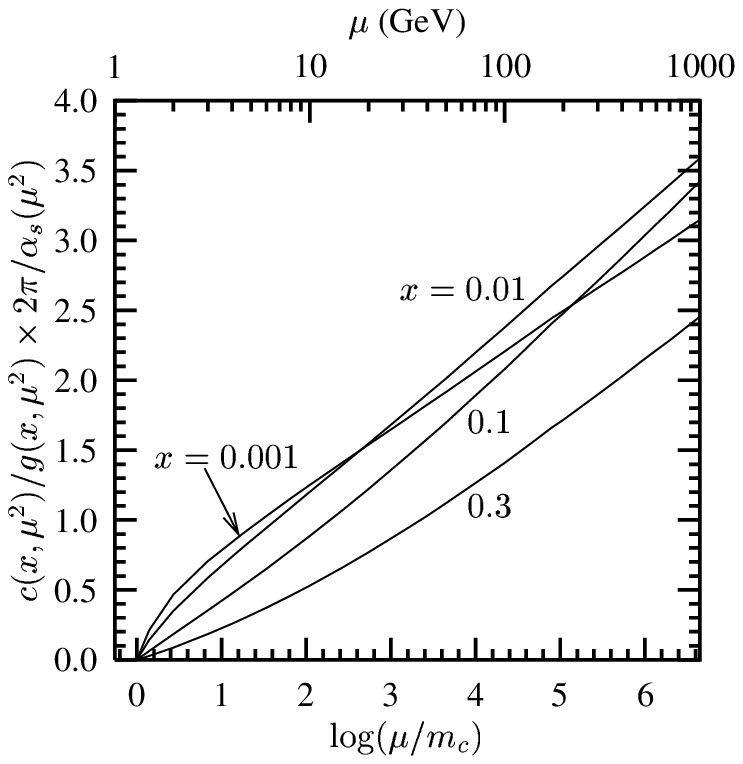}
\hspace*{1.5em}
\includegraphics{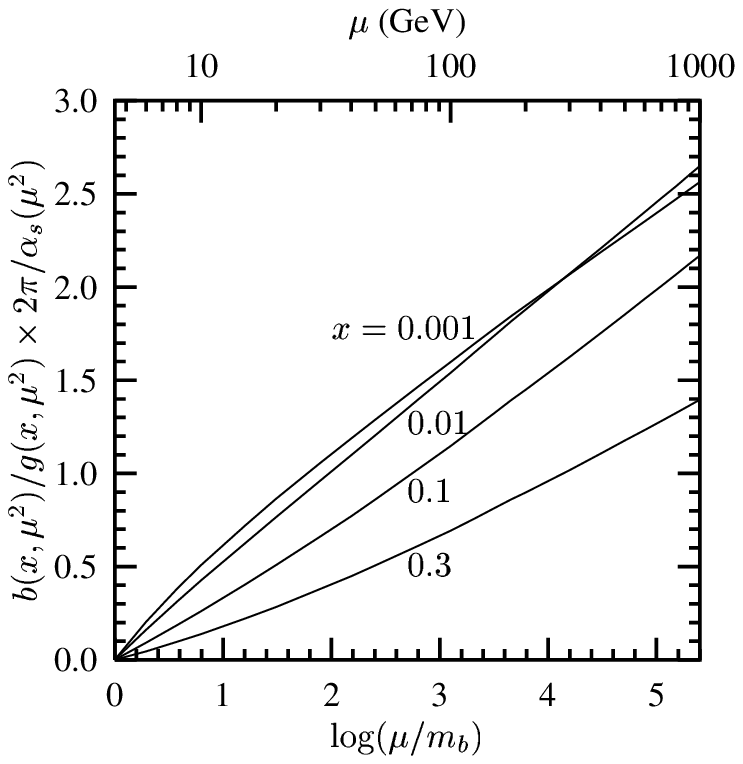}%
\vspace*{-1em}
\caption{The ratio of the $c$ and $b$ distribution functions to the
gluon distribution function, times $2\pi/\alpha_s(\mu^2)$, versus
$\ln\left( \mu/m_{Q}\right)$ for various fixed values of $x$.  The
curves are approximately linear, while the slope of the curves
saturates at about $0.5$ at small $x$, in agreement with the
approximation of Eq.~(\ref{P5_sullivan_0709eq_b2}).
\label{P5_sullivan_0709fig_bog}}
\end{center}
\vspace*{-2em}
\end{figure}

Since $g(x)$ grows as $x^{-n}$ ($n\approx 1.4$--$1.5$), at sufficiently
small $x$ Eq.~(\ref{P5_sullivan_0709eq_b}) may be rewritten as
\begin{equation}
\label{P5_sullivan_0709eq_b2}
\frac{Q(x,\mu^2)}{g(x,\mu^2)}\frac{2\pi}{\alpha_s(\mu^2)}
\approx 
\left( \frac{1}{n}-\frac{2}{n+1}+\frac{2}{n+2}\right) 
\ln \left( \frac{\mu}{m_Q}\right) 
\approx 0.5\ln \left( \frac{\mu}{m_Q}\right) \,.
\end{equation}
The approximation (\ref{P5_sullivan_0709eq_b2}) works remarkably well
over a wide range of $x$ and $\mu$.  In
Fig.~\ref{P5_sullivan_0709fig_bog} we present an updated version of
Fig.~5 of Ref.~\cite{Stelzer:1997ns}.  The ratios
$Q(x,\mu^2)/g(x,\mu^2)\times 2\pi /\alpha_{s}(\mu^2)$ are shown as
functions of $\mu$ for various fixed values of $x$, using the CTEQ5M1
parton distribution functions \cite{Lai:1999wy}.  The dependence on
$\ln(\mu/m_Q)$ is approximately linear indicating that, even at
next-to-leading order (NLO), $Q(x,\mu^2) \propto
[\alpha_s(\mu^2)/2\pi] \ln (\mu^2/m_Q^2) g(x,\mu^2)$.  Further, the
constant of proportionality saturates at $\sim 0.5$ for charm quarks
when $x\lesssim 0.1$, and for bottom quarks when $x\lesssim 0.05$.

If the heavy-quarks PDFs are directly proportional to the gluon PDF, it is
reasonable to expect that their uncertainties are approximately the same:
\begin{equation}
\frac{\delta Q(x,\mu^2)}{Q(x,\mu^2)}\approx 
\kappa_{Q}\frac{\delta g(x,\mu^2)}{g(x,\mu^2)}\; ,
\label{P5_sullivan_0709eqassume}
\end{equation}
where $\kappa_Q\approx 1$. We test this relation by estimating the PDF
uncertainties with the Hessian matrix method proposed J. Pumplin,
\textit{et al.} \cite{Pumplin:2001ct}. This paper contains 16 pairs of
PDF sets, where each pair corresponds to varying one independent
parameter $z_i$ in the PDF fit such that the $\chi^2$ of the fit
changes by $t^2=(5)^2=25$. We define the maximum positive and negative
errors on an operator $O$ by \cite{Pumplin:2001ct,Nadolsky:2001yg}
\begin{eqnarray}
\delta O_{+} & = & \frac{T}{t}\sqrt{\sum_{i=1}^{16}\Bigl( \max 
[\,O(z^{0}_{i}+t)-O(z_{i}^{0}),O(z^{0}_{i}-t)-O(z_{i}^{0}),0]\Bigr) ^2}
\;,\label{P5_sullivan_0709eqop1} \\
\delta O_{-} & = & \frac{T}{t}\sqrt{\sum_{i=1}^{16}\Bigl( \max 
[\,O(z_{i}^{0})-O(z^{0}_{i}+t),O(z_{i}^{0})-O(z^{0}_{i}-t),0]\Bigr) ^2}
\; ,\label{P5_sullivan_0709eqop2} 
\end{eqnarray}
where the ``tolerance'' $T$ is a scaling parameter that determines
the overall range of allowed variation of $\chi^2$ \cite{Pumplin:2001ct}.
In this study we use $T=10$.

\begin{figure}[tb]
\begin{center}
\includegraphics{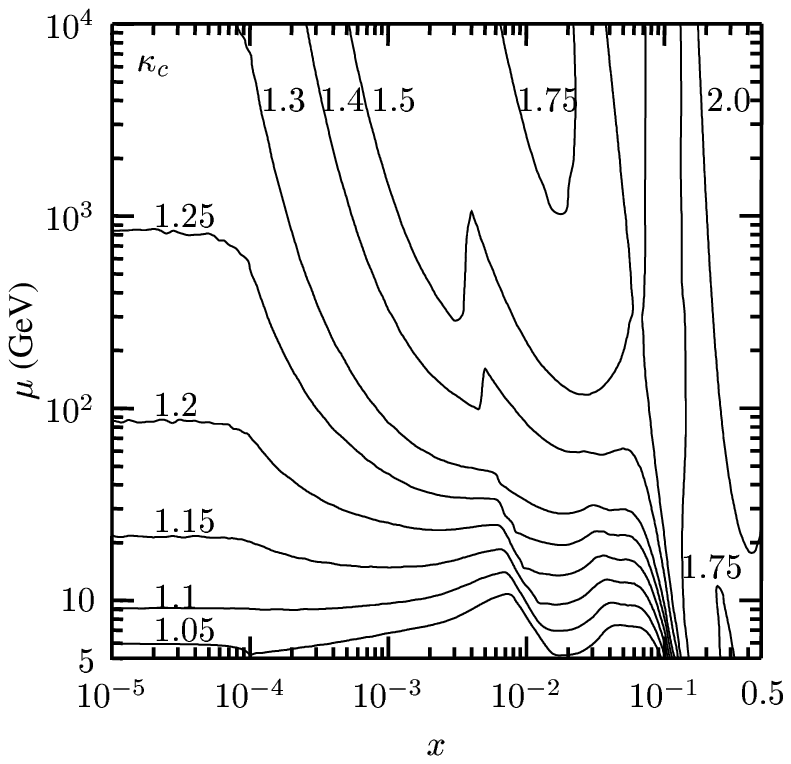}
\hspace*{1.5em}
\includegraphics{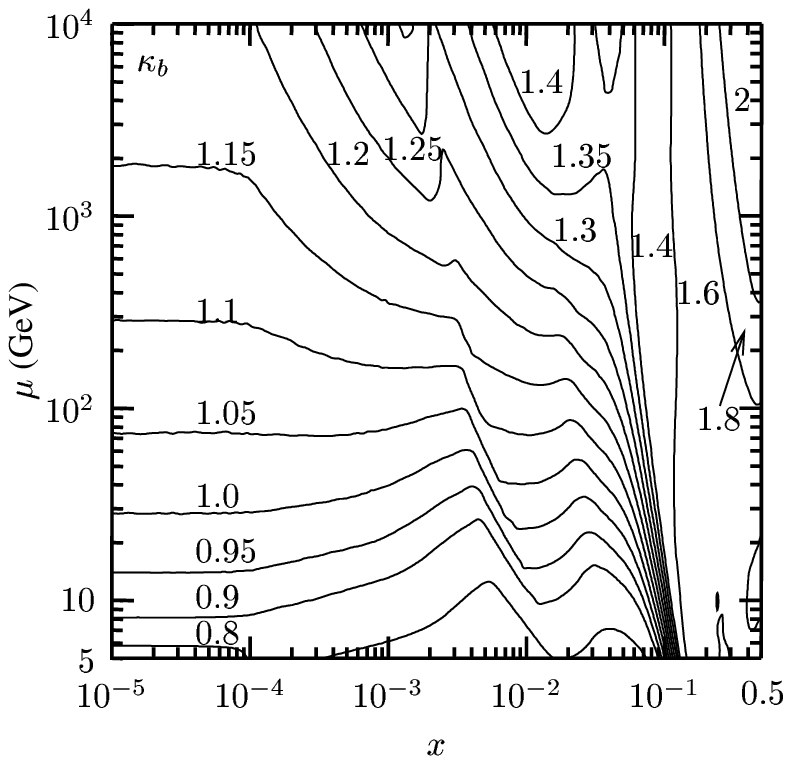}
\vspace*{-1em}
\caption{The ratios $\kappa_Q\equiv(\delta Q / Q)/(\delta g / g)$ for $c$ 
and $b$ quarks as a function of $x$ and $\mu$.}
\label{P5_sullivan_0709fig2}
\end{center}
\vspace*{-2em}
\end{figure}

Figure~\ref{P5_sullivan_0709fig2} shows the dependence of the ratios
$\kappa_c$ and $\kappa_b$ on $x$ and $\mu$.  We see that these ratios
are quite close to unity for $x<0.05$, but climb to about $2$ at
larger $x$.  The approximation (\ref{P5_sullivan_0709eqassume}) holds
better for $b$ quarks: $\kappa_b$ is less than $1.4$ at $x\leq 0.1$
and the whole range of $\mu$, while $\kappa_c$ can reach up to $1.75$
in this region.  The values $\kappa_Q$ are close to $1$ as
$x\rightarrow 0$, which supports the validity of
Eq.~(\ref{P5_sullivan_0709eq_b2}) in the small-$x$ region.

\begin{figure}[tb]
\begin{center}
\includegraphics{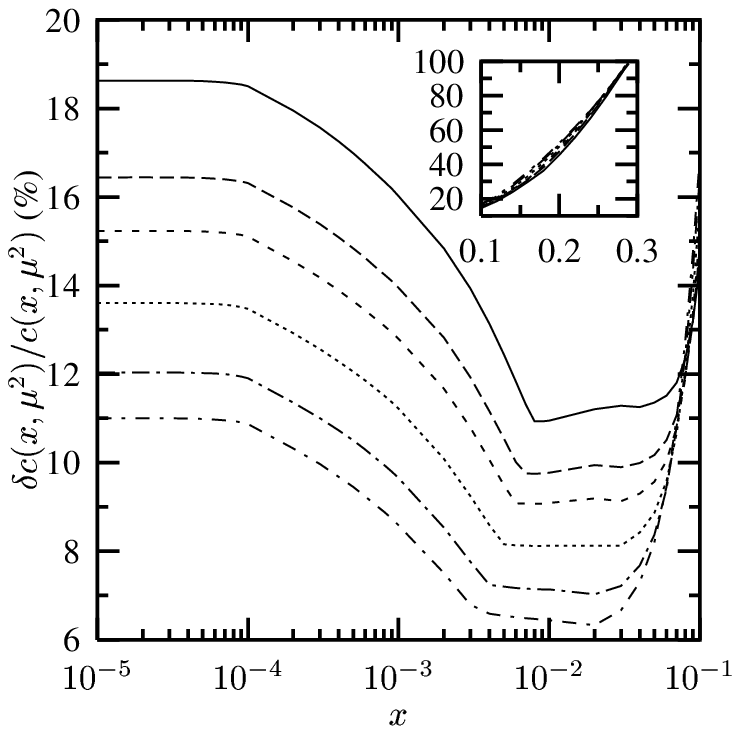}
\hspace*{1.5em}
\includegraphics{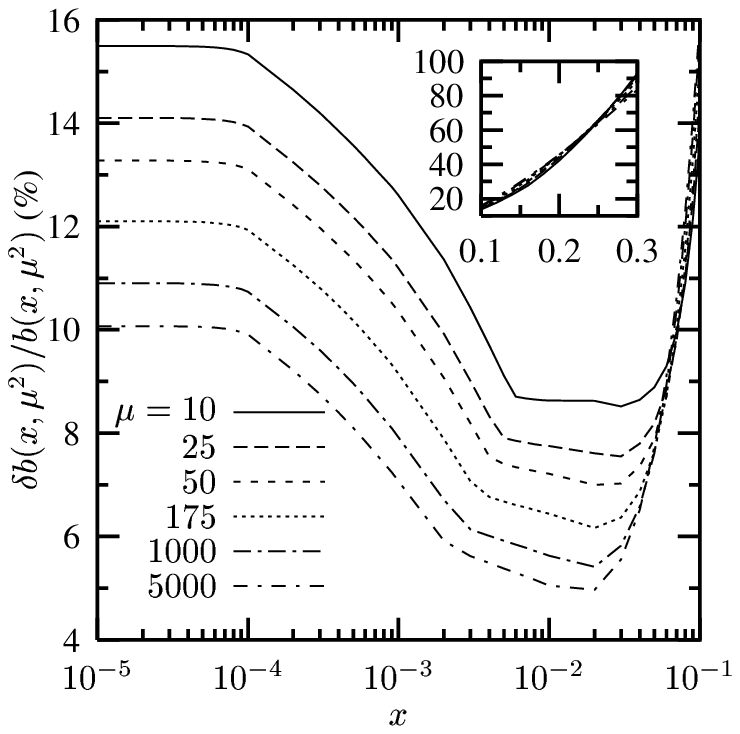}
\vspace*{-1em}
\caption{The uncertainties of the $c$ and $b$ distribution functions as 
a function of $x$ for various values of scale $\mu$.}
\label{P5_sullivan_0709fig3}
\end{center}
\vspace*{-3em}
\end{figure}

Figure~\ref{P5_sullivan_0709fig3} shows the relative uncertainties
$\delta c(x,\mu^2)/c(x,\mu^2)$ and $\delta b(x,\mu^2)/b(x,\mu^2)$ as
functions of $x$ at various values of $\mu$. Both uncertainties have
minima around $x\approx 0.01$, where the PDFs are best constrained by
the existing data. At small $x$, the uncertainties grow because the
small-$x$ region is covered only by DIS experiments, which do not
constrain well the gluon PDF. At $x\gtrsim 0.1$, the heavy-quark PDFs
become negligible compared to the valence quark PDFs, so that
$c(x,\mu^2)$ and $b(x,\mu^2)$ are practically unconstrained at
$x\gtrsim 0.3$.

Single-top-quark production via $t$-channel $W$-exchange probes the
$b$-quark PDF at $\mu\sim m_t\simeq 175$~GeV.  The range of $x$ probed
at the run II of the Tevatron for accepted events is $0.06$--$0.5$,
with the bulk of the cross section coming from the region around
$x\simeq 175/2000\simeq 0.09$.  Using this value of $x$ and
Fig.~\ref{P5_sullivan_0709fig3}b, we can estimate the uncertainty of
the total cross section at run II to be $\sim\pm 15\%$.  We have
explicitly re-calculated the NLO cross section in
Ref.~\cite{Stelzer:1997ns} and its uncertainty with the method
described above and $T=10$. The result is $2.12^{+0.32}_{-0.29}$~pb,
or $^{+15}_{-14}\%$, in excellent agreement with
Fig.~\ref{P5_sullivan_0709fig3}b.  For this choice of $T$, and using
CTEQ5M1 PDFs, the PDF uncertainties will be larger than all other
theoretical or experimental errors once $2$~fb$^{-1}$ of integrated
luminosity is accumulated.  One positive note is that
Fig.~\ref{P5_sullivan_0709fig3}b predicts and uncertainty of around
$\pm 7\%$ at the LHC, which will be comparable to experimental
systematics.

The approximation that the heavy-quark uncertainties are the same as the
gluon uncertainty is a good guess for smaller $x$ and $\mu$. We
calculate these uncertainties using a ``tolerance'' $T=10$ in
a Hessian matrix method~\cite{Pumplin:2001ct}. We show that the heavy-quark
uncertainties strongly depend on $x$ and relatively weakly on the scale
$\mu$. We calculate the PDF uncertainty for $t$-channel single-top-quark
production, and predict a cross section of $2.12^{+0.32}_{-0.29}$~pb
at run II of the Tevatron.
\vspace*{-1em}

\bibliographystyle{revtex}
\bibliography{P5_sullivan_0709}


\end{document}